\documentclass[12pt]{amsart}
\usepackage{graphicx}
\usepackage{verbatim}
\usepackage{amsmath}
\usepackage{amsfonts}
\usepackage{setspace}
\usepackage{url}

\begin{document}
\baselineskip=22pt

\newcommand{\logit}{{\rm logit}}

\newcommand{\sn}{\sum_{i=1}^n}
\newcommand{\snj}{\sum_{j=1}^n}
\newcommand{\snk}{\sum_{k=1}^\infty}
\newcommand{\intt}{\int_0^t}
\newcommand{\inti}{\int_0^\infty}
\newcommand{\bbeta}{\mbox{\boldmath $\beta$}}
\newcommand{\blambda}{\mbox{\boldmath $\lambda$}}
\newcommand{\bg}{\boldsymbol{g}}
\newcommand{\bX}{\boldsymbol{X}}
\newcommand{\bx}{\boldsymbol{x}}
\newcommand{\bm}{\boldsymbol{m}}
\newcommand{\bS}{\boldsymbol{S}}
\newcommand{\bSigma}{\mbox{\boldmath $\Sigma$}}
\newcommand{\bV}{\boldsymbol{V}}
\newcommand{\bU}{\boldsymbol{U}}
\newcommand{\bA}{\boldsymbol{A}}
\newcommand{\bW}{\boldsymbol{W}}
\newcommand{\bh}{\boldsymbol{h}}
\newcommand{\bu}{\boldsymbol{u}}
\newcommand{\bt}{\boldsymbol{t}}
\newcommand{\bI}{\boldsymbol{I}}
\newcommand{\bvarepsilon}{\mbox{\boldmath $\varepsilon$}}
\newcommand{\bhbeta}{\mbox{\boldmath $\widehat \beta$}}
\newcommand{\bD}{\boldsymbol{D}}
\newcommand{\bM}{\boldsymbol{M}}

\newtheorem{thm}{Theorem}[section]
\newtheorem{cor}[thm]{Corollary}
\newtheorem{lem}{Lemma}
\newtheorem{prop}[thm]{Proposition}
\newtheorem{defn}[thm]{Definition}
\newtheorem{rem}[thm]{Remark}
\newtheorem{eg}{example}[section]
\newcommand{\norm}[1]{\left\Vert#1\right\Vert}
\newcommand{\abs}[1]{\left\vert#1\right\vert}
\newcommand{\set}[1]{\left\{#1\right\}}
\newcommand{\Real}{\mathbb R}
\newcommand{\eps}{\varepsilon}
\newcommand{\To}{\rightarrow}
\newcommand{\BX}{\mathbf{B}(X)}
\newcommand{\A}{\mathcal{A}}
\newcommand{\goto}{\longrightarrow}
\newcommand{\gotoinp}{\overset{p_r}{\goto}}
\newcommand{\gotoind}{\overset{d}{\goto}}

$ $

\vspace{-8mm}

\begin{flushright}
{\bf \LARGE An Empirical Likelihood Approach to Nonparametric
Covariate Adjustment in Randomized Clinical Trials}
\end{flushright}

\begin{flushleft}
\bf Xiaoru Wu and Zhiliang Ying
\end{flushleft}
\vspace{-5mm}

\noindent
---------------------------------------------------------------------------------------------------------------
\noindent Covariate adjustment is an important tool in the
analysis of randomized clinical trials and observational studies.
It can be used to increase efficiency and thus power, and to
reduce possible bias. While most statistical tests in randomized
clinical trials are nonparametric in nature, approaches for
covariate adjustment typically rely on specific regression models,
such as the linear model for a continuous outcome, the
logistic regression model for a dichotomous outcome and the Cox model for
survival time. Several recent efforts have focused on model-free
covariate adjustment. This paper makes use of the empirical
likelihood method and proposes a nonparametric approach to
covariate adjustment. A major advantage of the new approach is
that it automatically utilizes covariate information in an optimal
way without fitting nonparametric regression. The usual asymptotic
properties, including the Wilks-type result of convergence to a
$\chi^2$ distribution for the empirical likelihood ratio based
test, and asymptotic normality for the corresponding maximum
empirical likelihood estimator, are established. It is also shown
that the resulting test is asymptotically most powerful and that
the estimator for the treatment effect achieves the semiparametric
efficiency bound. The new method is applied to the Global Use of
Strategies to Open Occluded Coronary Arteries (GUSTO)-I trial.
Extensive simulations are conducted, validating the theoretical
findings.

\vspace{2mm}

\noindent
---------------------------------------------------------------------------------------------------------------
\noindent  Xiaoru Wu is PhD candidate, Department of Statistics,
Columbia University, New York, NY 10027 (Email: {\it
xw2144@columbia.edu}) and Zhiliang Ying is Professor, Department of
Statistics, Columbia University, New York, NY 10027 (Email: {\it
zying@stat.columbia.edu}). The authors thank the Virtual
Coordinating Center for Global Collaborative Cardiovascular Research
(VIGOUR) Leaders for the use of GUSTO-I data. This research was
supported by grants from the National Institutes of Health and the
National Science Foundation.
\newpage

\section{INTRODUCTION}

Testing for the statistical significance of treatment differences is a
key element in the analysis of randomized clinical trials. In
its simplest form, patients are randomly allocated to either a
treatment or control group and their responses are recorded.
Many statistical methods are available for testing whether there is
convincing evidence that a treatment difference exists between the two
groups; cf. Pocock (1983) and Friedman, Furberg and DeMets (1998).
In addition to treatment allocation and outcome values, baseline
covariate information is often collected in such clinical studies. Classical
analysis of covariance (ANCOVA) and other regression model-based
tests may be used to handle covariate adjustment; cf. Scheffe
(1959), Simon (1984), McCullagh and Nelder (1989) and Rutter and
Elashoff (1994). When properly used, covariate adjustment can
increase efficiency and, in the case of an observational study,
reduce bias (Armitage 1981).

Due to randomization, most two-sample (multi-sample if more than two
treatment groups are involved) tests are valid without any
parametric assumption. Therefore, these tests are nonparametric in
nature, a feature of great importance in a clinical trial. Standard
methods for covariate adjustment, however, require that a specific
regression model be assumed; see, for example, Piantadosi (2005,
Chapter 17).

Adjusting for covariates without assuming a regression model has
been studied by Koch (1998), Tsiatis, Davidian, Zhang and Lu (2008)
among others. In particular, Koch (1998) proposed a weighted least
squares method to include covariate information for estimating the
treatment difference. This method always leads to a variance
reduction, thus an increase in power.  By appealing to
semiparametric efficiency theory, Tsiatis et al. (2008) developed a
general approach to covariate adjustment that circumvents modeling
the covariate-outcome relationship.  Their approach allows for
nonlinear terms in relating the auxiliary covariates to the outcome
variable, thereby further reducing the variability. They showed that
the method is semiparametrically efficient by deriving the
semiparametric information bound and by showing the bound is
attained with their approach.

An essential ingredient in the approach by Tsiatis et al. (2008) is
the use of the independence of treatment allocation and baseline
covariates to construct equations associated.
These equations can be viewed as constraints
that, when properly utilized, may lead to further reduction in
variability of the outcome variable. How to optimally use these
constraints is therefore crucial for efficiency improvement.

Empirical likelihood (Owen 1988) is a general method for efficiently
utilizing constraints or estimating equations. Specifically, it
maximizes the nonparametric likelihood (Kiefer and Wolfowitz 1956)
subject to certain constraints that are specific to the problem of
interest. It can be used to obtain empirical likelihood ratio tests
as well as confidence intervals. Examples include testing and
interval estimation for population means and for regression
coefficients. Qin and Lawless (1994) showed that the constraints can
be used more liberally in the sense that the number of constraints
may exceed the number of parameters of interest. They also showed
that the empirical likelihood utilizes the information in the
constraints in an optimal way.

Because baseline covariate information for a randomized clinical
trial generates constraints, it is natural to consider the empirical
likelihood as a means to improve efficiency for the primary problem
of testing and estimating treatment difference. To that end, this
paper proposes a general approach to covariate adjustment by making
use of the empirical likelihood and suitably choosing
constraints. The new approach does not require any model assumption
on the relationship between the outcome variable and baseline
covariates. It is shown that such an empirical likelihood based
method automatically results in efficiency improvement. For testing,
it is asymptotically most powerful; for estimation, it achieves the
semiparametric information bound.

The rest of the paper is organized as follows. In Section 2 we
introduce some notation and briefly discuss existing model-based
methods.  We apply the empirical likelihood method for covariate
adjustment and extend it to inference with growing number of
constraints in Section 3.  The design and results of simulation
studies are described in Section 4. In Section 5, the method is
applied to a study of acute myocardial infarction. Some concluding
remarks are given in Section 6.

\vspace{-8mm}

\section{NOTATION AND MODEL SPECIFICATION}

In a $(K+1)$-arm ($K\geq1$) randomized clinical trial, for
subject $i$, let $Y_i$, $Z_i$ and $\bX_i$ denote the outcome,
treatment allocation and available auxiliary baseline covariates,
respectively. Assume that $(Y_i, Z_i, \bX_i)$, $i=1,\ldots, n$, are
independent and identically distributed (i.i.d.) and that the random
allocation probabilities $\pi_k=P(Z=k)$, $k=0,\ldots,K$, where
$\sum_{k=0}^K\pi_k=1$, are known.

Throughout, $G^k$ denotes the conditional distribution
of the outcome variable $Y$ given treatment allocation $Z=k$,
$k=0,\ldots,K$. Then the usual null hypothesis of no treatment
difference is given by\vspace{-1mm}
\begin{equation*} H_0:
G^0=G^1=\ldots=G^K.
\end{equation*}\vspace{-1mm}
Note that there is no assumption on the form of $\{G^k,
k=0,\ldots, K\}$.

To study treatment effects, one may choose certain contrasts
among the treatment groups in terms of their population
characteristics, for example, the difference in mean outcomes between
two treatment groups. Following Zhang et al. (2008), the treatment
effect can be identified by considering\vspace{-3mm}
\begin{equation}
\beta_1=E(Y|Z=0),\quad \beta_2=E(Y|Z=1)-E(Y|Z=0),\vspace{-3mm}
\end{equation}
or equivalently, by formulating\vspace{-3mm}
\begin{subequations}
\begin{align}\label{mean2gr} E(Y|Z)=\beta_1+\beta_2 Z.\vspace{-4mm}
\end{align}
Clearly, such an approach does not require model assumption on the
underlying distribution functions $G^k$, $k=0, \dots, K$. If there
are more than two treatment groups, equation (\ref{mean2gr})
becomes\vspace{-5mm}
\begin{align}\label{meanKgr}
E(Y|Z)=\beta_1+\beta_2 1_{(Z=1)}+\ldots+\beta_{K+1}
1_{(Z=K)},\vspace{-3mm}
\end{align}\end{subequations}where $1_{(\cdot)}$ is the indicator function and  $\beta_{k+1}$ represents the difference in mean
outcome between group $k$ and group $0$. For a binary outcome, an
alternative formulation is via the log-odds ratios:
\begin{equation}
\logit\{P(Y=1|Z)\}=\log \left\{\frac {P(Y=1|Z)}
{P(Y=0|Z)}\right\}=\beta_1+\beta_2 1_{(Z=1)}+\ldots+\beta_{K+1}
1_{(Z=K)}.\vspace{-3mm}
\end{equation}
Under this formulation, testing the null hypothesis of no treatment
difference is tantamount to testing $H_0:
\beta_2=\ldots=\beta_{K+1}=0$, and estimating the treatment effect
is tantamount to estimating values of the $\beta_k, k=2,\ldots,
K+1$. For notational convenience, we use $\bbeta$ to denote the
parameter vector $(\beta_1, \dots, \beta_{K+1})^T$.

Besides the outcome variable and treatment assignment, relevant
baseline covariates, which may comprise patients' demographic
information, medical history, lifestyle measurements, etc., may be
recorded as well. Their association with and impact on the outcome
variable can then be explored for efficiency gains in testing and
estimation of treatment effects. A common approach to
adjusting for covariates is to postulate a certain regression model,
which gives treatment comparisons conditional on values of
the covariates. It is well known that treatment effects may have
different interpretations in conditional and unconditional (on
covariate value) models. Indeed, except for linear and exponential
regression models, the conditional and unconditional approaches
generally lead to different parameter values for the treatment
effect. We refer to Gail (1984) for a comprehensive discussion on
the subject.

Since the unconditional treatment effect is of primary interest
here, it is natural for us to avoid any modeling of the relationship
between the outcome variable and baseline covariates. Yet it is also
desirable that we make best use of the information in the covariates
to improve efficiency. To this end, we explore the empirical
likelihood methodology to develop a model-free approach to covariate
adjustment. We demonstrate that such an approach is natural for
nonparametric covariate adjustment and optimal in terms of
efficient use of available information.

\vspace{-3mm}
\section{EMPIRICAL LIKELIHOOD BASED METHODS FOR NONPARAMETRIC COVARIATE ADJUSTMENT}

Being first implicitly used in Thomas and Grunkemeier (1975),
empirical likelihood was developed into a general methodology by
Owen (1988, 1990).  Given $(Y_i, Z_i, \bX_i)$, $i=1,\ldots, n$,
assumed to be independent with a common cumulative
distribution function (CDF) $F_0$, the empirical likelihood function
is a nonparametric likelihood function of the CDF $F$\vspace{-3mm}
\begin{equation}\label{EL}
L(F)=\prod_{i=1}^{n}dF(y_i,z_i,\bx_i)=\prod_{i=1}^{n}
p_i,\vspace{-3mm}
\end{equation}
where $(y_i,z_i,\bx_i)$ is the observed value of $(Y_i, Z_i,
\bX_i)$, $p_i=dF(y_i,z_i,\bx_i)=P(Y_i=y_i, Z_i=z_i, \bX_i=\bx_i)$,
$i=1,\ldots, n$. Without additional constraints (other than $p_i\geq
0$ and $\sum_{i=1}^n p_i=1$), it is well known that the empirical
distribution function is the nonparametric maximum likelihood
estimate of $F_0$.

This section is devoted to the development of an empirical
likelihood based method for nonparametric covariate adjustment
arising from a typical randomized clinical trial. Subsection 3.1
develops an empirical likelihood ratio based test and establishes
its asymptotic properties. The subsequent subsection deals with the
dual problem of estimating treatment effects via maximizing the empirical
likelihood with the number of constraints exceeding the number of
parameters. Subsection 3.3 extends the results of 3.1 and 3.2 to the
situation in which the number of constraints increases with the
sample size. Asymptotic normality and Wilks type $\chi^2$
approximation as well as asymptotic efficiency are established for
all the cases under suitable regularity conditions.

\noindent 3.1  \textbf{Testing Treatment Differences}\\
\noindent Empirical likelihood methodology for inference is based on
maximizing the nonparametric likelihood (\ref{EL}) subject to
appropriately formulated and problem-specific constraints. For the
two-arm randomized clinical trial specified by (\ref{mean2gr}), the
constraints are generated by\vspace{-4mm}
\begin{subequations}\label{meangr}
\begin{align}\label{m}
\bm(\bbeta;Y,Z)=(1,Z)^T(Y-\beta_1-\beta_2Z).\end{align} For general
$K$ specified by (\ref{meanKgr}), it becomes
\begin{align}\label{mK}\bm(\bbeta;Y,Z)=(1, 1_{(Z=1)},\ldots,
1_{(Z=K)})^T(Y-\beta_1-\beta_21_{(Z=1)}-\ldots-\beta_{(K+1)}1_{(Z=K)})
.\end{align}
\end{subequations}
 The zero-mean property of $\bm(\bbeta;Y,Z)$ uniquely
determines the value of $\bbeta$ and can be used to obtain
estimators through the sample-generated estimating equations. The
resulting inference involves only the $Y_i$ and $Z_i$.

The availability of the baseline covariates $\bX_i$ should enable us
to obtain additional estimating equations, thereby additional
constraints. Indeed, Davidian et al. (2005) and Leon et al. (2003)
found that the following form gives a general family of estimating
equations:\vspace{-3mm}
\begin{equation}\label{unbias}
\sum_{k=0}^K(1_{(Z=k)}-\pi_k)h_k(\bX),\vspace{-3mm}
\end{equation}
where $h_k$, $k=0, 1, \dots, K$ are arbitrary functions.  The
independence of $Z$ and $\bX$ guarantees the zero-mean property of
the resulting estimating equations.

It is clear now that the number of zero-mean estimating equations as
provided by (\ref{meangr}) and (\ref{unbias}) exceeds the number of
parameters which specify the treatment effect. In fact, the number
of possible equations that can be generated from (\ref{unbias}) can
be unlimited when the baseline covariates $\bX$ are continuous.
Suppose we fix the choice of $h_k$ and consider how to make use of
them for efficiency improvement. For notational simplicity, we use
$\bg_r(\bbeta;Y,Z,\bX)$ to denote an $r$-vector of the resultant
estimating equations that include both (\ref{meangr}) and
(\ref{unbias}). Here $r\geq 2$ in the two-sample case and $r\ge K+1$
for the general $(K+1)$-sample case.

It is well known that the empirical likelihood approach links
together the inference of certain parameters and the available
estimating equations to form a constrained optimization problem.
With constraints given by $\bg_r$, it maximizes $L(F)$ in (\ref{EL})
subject to the following constraints:\vspace{-4mm}
\begin{equation}\label{constraints} p_i\geq
0,\quad \sum_{i=1}^np_i=1,\quad
\sum_{i=1}^np_i\bg_r(\bbeta;Y_i,Z_i,\bX_i)=0.\vspace{-4mm}
\end{equation}
This optimization problem has a unique maximizer provided that 0 is
inside the convex hull of $\{\bg_r(\bbeta;y_i,z_i,\bx_i),
i=1,\ldots,n\}$ for a given $\bbeta$ (Owen 2001). By applying the
Lagrange multiplier argument (Lang 1987), we can easily get
$p_i=\{n[1+\widehat{\blambda}^T(\bbeta)\bg_r(\bbeta;y_i,z_i,\bx_i)]\}^{-1}$,
where $\widehat{\blambda}$, which is a function of $\bbeta$, is the
solution to\vspace{-3mm}
\begin{equation}\label{lam}
\frac{1}{n}\sum_{i=1}^{n}\frac{\bg_r(\bbeta;y_i,z_i,\bx_i)}{1+\widehat{\blambda}^T
(\bbeta)\bg_r(\bbeta;y_i,z_i,\bx_i)}=0.\vspace{-3mm}
\end{equation}
Therefore, the resulting profile empirical log-likelihood, as a
function of $\bbeta$, takes form\vspace{-3mm}
\begin{equation}\label{loglik} l_E(\bbeta) =
\sum_{i=1}^{n}\log\Big[1+\widehat{\blambda}^T(\bbeta)\bg_r(\bbeta;y_i,z_i,\bx_i)\Big].\vspace{-3mm}
\end{equation}

\begin{thm}\label{fixthm1} Let $\bbeta^T=(\bbeta_1^T, \bbeta_2^T)$, where
$\bbeta_1$ and $\bbeta_2$ are $q_1$- and $q_2$-vectors.
Define\vspace{-3mm}
\begin{equation}\label{TE}
T_E=2l_E(\widehat{\bbeta}_{10},0)-2l_E(\widehat{\bbeta}),\vspace{-3mm}
\end{equation}
the logarithmic empirical profile likelihood ratio for testing
$\widetilde{H_0}: \bbeta_2=0$, where $\widehat{\bbeta}_{10}$
minimizes $l_E(\bbeta_1,0)$ with respect to $\bbeta_1$ and
$\widehat{\bbeta}$ minimizes $l_E(\bbeta)$. Then, under some mild
regularity conditions, $T_E$ converges to $\chi^2_{(q_2)}$ in
distribution under $\widetilde{H_0}$.\vspace{-5mm}
\end{thm}
Theorem~\ref{fixthm1} is a direct adaptation of Corollary 5 in Qin
and Lawless (1994).  It enables us to get the $p$-value in testing
the null hypothesis of no treatment difference and to invert the
test to obtain the confidence limits.  A numerical way to find
$\widehat{\bbeta}$, and similarly for $\widehat{\bbeta}_{10}$, is to
use a two-stage Newton algorithm. We first specify an initial
value $\bbeta^{(0)}$ for $\bbeta$ and solve (\ref{lam}) to obtain
$\widehat{\blambda}(\bbeta^{(0)})$. Next, we fix
$\widehat{\blambda}(\bbeta)$ in (\ref{loglik}) at
$\widehat{\blambda}(\bbeta^{(0)})$ and minimize (\ref{loglik}) over
$\bbeta$ to obtain a new value $\bbeta^{(1)}$.  We iterate the
process until convergence.

 From Qin and Lawless (1994), it follows that the empirical
likelihood ratio test incorporating covariate information through
constraints $\bg_r(\bbeta;Y, Z, \bX)$ is always more powerful than
the one with $\bm(\bbeta;Y, Z)$ only. Moreover, the more
constraints we put into $\bg_r$,  the more powerful the test
becomes. Because the net effect of the empirical likelihood method
with more constraints than parameters is an optimal linear
combination of the constraints, choice of additional constraints
should therefore be made to avoid redundancy. However, it is not
necessary to model the relationship between the covariates and the
outcome, as is evident from equation (\ref{unbias}); this is a
very desirable feature with important practical implications.

For a binary outcome variable, if we are interested in using the
log-odds ratio, then we can replace (\ref{mK}) with
$$\bm(\bbeta;Y,Z)=(1,
1_{(Z=1)},\ldots,1_{(Z=K)})^T[Y-\phi(\beta_1+\beta_21_{(Z=1)}+
\ldots+\beta_{(K+1)}1_{(Z=K)})],$$ where
$\phi(\cdot)=\exp(\cdot)/[1+\exp(\cdot)]$ is the logistic function.
We can then follow the same steps to
construct the empirical likelihood ratio test. As before, the large
sample properties given by Theorem 3.1 continue to hold.

\noindent 3.2 \textbf{Maximum Empirical Likelihood Estimate of
Treatment Effect}\\ \noindent Without adjusting for baseline
covariates, the number of estimating equations, derived from the
score functions, equals the number of parameters. Solving
equations $\sum_{i=1}^nm(\bbeta;Y_i,Z_i)=0$ gives us the M-estimator
for $\bbeta$, which is known to be consistent and asymptotically
normal (Huber 1981). With covariate adjustment, we have additional
estimating equations containing auxiliary information through
(\ref{unbias}).  Since the number of all available estimating
equations $r$ exceeds the number of parameters $q=q_1+q_2$, we
cannot obtain the estimators simply by finding zeros of those
estimating equations. One way to handle overly constrained problem
is to form $q$-dimensional linear combinations of all available
estimating equations so that the resulting set of equations has a
unique solution.  One can further evaluate the limiting covariance
matrix of the estimator to identify the optimal choice of such
linear combinations;  cf. Goldambe and Heyde (1987). Because the
empirical likelihood method with overly constrained estimating
equations can result in the optimal combination (Qin and Lawless
1994), it provides a nature alternative. The following result
follows directly from Qin and Lawless (1994).
\nopagebreak\vspace{-5mm}
\begin{thm}\label{fixthm2}
Let $\bD_r = E[\partial \bg_r(\bbeta_0)/\partial\bbeta^T]$ and
$\bSigma_r = E(\bg_r\bg_r^T)$. Then, under certain regularity
conditions, we have\nopagebreak\vspace{-3mm}\begin{equation}
n^{1/2}(\widehat{\bbeta}-\bbeta_0)\To
N\Big(0,(\bD_r^T\bSigma_r^{-1}\bD_r)^{-1}\Big),\vspace{-3mm}\end{equation}
where $\widehat{\bbeta}$ is the maximum empirical likelihood
estimate (MELE).\end{thm}

The theorem above allows us to construct Wald-type confidence
intervals using the robust variance estimate.  From Corollary 2 of
Qin and Lawless (1994), it follows that $\widehat{\bbeta}$ has the
smallest asymptotic variance among all the $q$-dimensional linear
combinations of $\bg_r(\bbeta;Y,Z,\bX)$. In particular, when $r=q$,
the maximum empirical likelihood estimator $\widehat{\bbeta}$ will
be asymptotically equivalent to the M-estimator. Furthermore,
Corollary 1 of Qin and Lawless (1994) ensures that the more
constraints being put into the optimization problem, the more
precision one can achieve.

As an example, consider again a two-arm clinical
trial with a binary outcome variable and a continuous covariate $X$,
and suppose the log-odds ratio is of interest. We can incorporate both
linear and quadratic terms of $X$ by using
constraints
$$\bg_r(\bbeta;Y,Z,X)=\Big((1,
Z)[Y-\phi(\beta_1+\beta_2Z)],(Z-\pi_1),
(Z-\pi_1)X,(Z-\pi_1)X^2\Big)^T.$$  The resulting estimator will be
more efficient than the M-estimator from $(1,
Z)^T[Y-\phi(\beta_1+\beta_2Z)]$. Note that, for regression model
based covariate adjustment, Robinson and Jewell (1991)
demonstrated that including predictive covariates in the logit
will always result in a loss of precision. In contrast, for our
empirical likelihood approach, including predictive covariates in
the constraints will never lead to an increase in the asymptotic
variance. The fact that incoporating additional estimating
equations always improves efficiency makes the empirical
likelihood approach advantageous and convenient.

\noindent 3.3 \textbf{Empirical Likelihood With Growing Number of
Constraints}\\ \noindent Since we can achieve more precision by
increasing the number of constraints, it is intuitive that semiparametric
efficiency may be attained when the number of constrains grows with
the sample size. In this connection, we consider in this subsection
the empirical likelihood based covariate adjustment when the number
of constraints grows to infinity as $n\to \infty$. Note here that
the dimension of $\bbeta$, which is of primary concern, remains fixed.

Suppose besides the $q$-dimensional score $\bm(\bbeta;Y,Z)$, the
auxiliary information is contained in an $r_n$-vector of estimating
equations $\bg^*_{r_n}(\bbeta)=(\bm^T(\bbeta;Y,Z),\bV_{n}^T)^T$.
Instead of a fixed number $r$, $r_n$ here will grow to infinity with
$n$ at a certain rate.  The $j^{th}$ component of $\bV_{n}$ has the
form $(1_{(Z=k)}-\pi_k)h_j(\bX)$ for $j=1,\ldots,r_n-q$, where $h_j$
is a real-valued function.  The following conditions will be used.

\noindent (C1) There exists a non-random $(r_n-q)\times(r_n-q)$
matrix $\bW_n$ such that (i)-(iii) below are satisfied for
$\bg_{r_n}(\bbeta)= (\bm^T(\bbeta;Y,Z),(\bW_nV_{n})^T)^T$.\\
(i) Components of $\bg_{n,i}$, $i=1,\ldots, n$, are uniformly
bounded by a finite constant $M > 0$,  where $\bg_{n,i}(\bbeta)=\bg_{r_n}(\bbeta;Y_i, Z_i, \bX_i)$.\\
(ii) Eigenvalues of
$\bSigma_{n,g}=E(\bg_{r_n}(\bbeta_0)\bg_{r_n}^T(\bbeta_0))$ are
bounded away from zero and infinity.\\
(iii) There exists a $q\times (r_n-q)$ non-random matrix $\bA_n$
such that $$\bA_n\bW_n\bV_{n}\To
\sum_{k=0}^K(1_{(Z=k)}-\pi_k)E(\bm(\bbeta;Y,Z)|Z=k,\bX)\quad
\text{in
$\mathbb{L}^2$}.$$ \\
(C2) The growth rate of ${r_n}$ is limited to ${r_n^3}=o(n)$.\\
(C3) Matrix
$\widetilde{\bSigma}=E(\widetilde{\bm}\widetilde{\bm}^T)$ is
positive definite, where
$$\widetilde{\bm}=\bm(\bbeta;Y,Z)-\sum_{k=0}^K(1_{(Z=k)}-\pi_k)E(\bm(\bbeta;Y,Z)|Z=k,\bX).$$

\begin{thm}\label{thm1} Let $\widehat{\bbeta}_n$ be the maximum empirical likelihood
estimate based on constraints $\bg^*_{r_n}(\bbeta)$ and $\bD_m =
E(\partial \bm(\bbeta_0)/\partial\bbeta^T)$. Then, under Conditions
C1-C3,\vspace{-3mm}\begin{equation}\label{asyvar}
n^{1/2}(\widehat{\bbeta}_n-\bbeta_0)\To
N\Big(0,(\bD_m^T\widetilde{\bSigma}^{-1}\bD_m)^{-1}\Big).\end{equation}
\end{thm}

Minimizing the asymptotic variance of the M-estimator from the class
of arbitrary $q$-dimensional unbiased estimating equations, Zhang et
al. (2008) derived the semiparametric efficiency bound for the
estimators of treatment effect.  From Zhang et al. (2008) and
Theorem~\ref{thm1}, we have the following result.

\begin{cor}\label{cor3}The limiting variance-covariance matrix,
$(\bD_m^T\widetilde{\bSigma}^{-1}\bD_m)^{-1}$, achieves the
semiparametric efficiency bound, i.e., $\widehat{\bbeta}_n$ in
Theorem~\ref{thm1} is asymptotically efficient.\end{cor}

In practice, in order to construct the Wald type confidence interval
for $\bbeta_0$, we need to estimate the asymptotic variance
expressed in (\ref{asyvar}).  Let
$\overline{\bg}_n(\bbeta)=n^{-1}\sum_{i=1}^n \bg_{n,i}(\bbeta)$,
$\bS_n(\bbeta)=n^{-1}\sum_{i=1}^n\bg_{n,i}(\bbeta)\bg_{n,i}^T(\bbeta)$
and $\hat{\bD}(\bbeta) = \partial
\overline{\bg}_n(\bbeta)/\partial\bbeta^T$. Theorem
\ref{varestimate} below shows that a consistent estimate of the
limiting variance-covariance matrix of
$n^{1/2}(\widehat{\bbeta}_n-\bbeta_0)$ is
$[\hat{\bD}(\hat{\bbeta}_n)\bS_n^{-1}(\widehat{\bbeta}_n)\hat{\bD}(\hat{\bbeta}_n)]^{-1}$.

\begin{thm}\label{varestimate} Under Conditions C1-C3,
$\norm{\hat{\bD}(\hat{\bbeta}_n)\bS_n^{-1}(\widehat{\bbeta}_n)\hat{\bD}(\hat{\bbeta}_n)
-\bD_m^T\widetilde{\bSigma}^{-1}\bD_m} = o_p(1)$.
\end{thm}

Throughout, $\norm{\cdot}$ is used to denote the
Euclidean norm.
 Theorem~\ref{thm1} states that the listed conditions are sufficient
to ensure standard asymptotic properties of the MELE. Moreover,
Corollary~\ref{cor3} states that when the number of constraints
grows to infinity at a certain rate, the MELE achieves the
semiparametric efficiency as derived in Zhang (2008). In
Theorem~\ref{thm1}, $\bg_{r_n}$ is essentially a linear
transformation of $\bg_{r_n}^*$.  Since a linear transformation does
not change the constraints, the estimator using $\bg_{r_n}$ will be
the same as that using $\bg_{r_n}^*$.  The fact that the MELE will
not be affected by a linear transformation of the constraints
greatly facilitates the applicability of the empirical likelihood
approach because we can just throw in all the constraints we have
without forming the appropriate combination of them.  For example,
$E[\bg^*_{r_n}(\bg^*_{r_n})^T]$ might be ill conditioned but we can
still use it as long as there exists a $\bW_n$ such that the
corresponding $\bSigma_{n,g}$ is better conditioned. For this
reason, we will not distinguish among linear transformations of
constraints in the following discussion.

Theorem~\ref{thm1} holds for a general $q$-dimensional score $\bm$ as
long as some regularity conditions in the case of fixed number of
constraints (Qin and Lawless 1994) are satisfied, including
$E\Big(\partial \bm(\bbeta;Y,Z)/\partial\bbeta^T\Big)$ is of full
rank $p$, $\norm{\partial \bm(\bbeta;Y,Z)/\partial\bbeta^T}$ and
$\norm{\partial^2 \bm(\bbeta;Y,Z)/\partial\bbeta\partial\bbeta^T}$
can be bounded by some integrable function in a neighborhood of
$\bbeta_0$ and $\partial \bm(\bbeta;Y,Z)/\partial\bbeta$ and
$\partial^2 \bm(\bbeta;Y,Z)/\partial\bbeta\partial\bbeta^T$ are
continuous in this neighborhood.

Condition C2 imposes an upper bound on the growth rate of the number
of constraints at which a well-behaved MELE can be obtained. In
practice, the number of constraints need not be large.  In fact, we
find that additional gain by including an extra constraint
diminishes quickly, due to the optimal use of constraints by the
empirical likelihood method.  It is important to note that the
asymptotic normality and efficiency are not affected by the choice
of $r_n$, as long as it satisfies C2.  It is certainly of
theoretical interest to find the sharp upper bound for $r_n$ to grow
such that the resulting estimate is still asymptotically normal and
efficient.  But we will not get into this complication here since
finding the optimal rate is not our main concern. If we knew the
conditional expectations in Condition C3, the optimal estimating
equations $\widetilde{\bm}$ would be the constraints that lead to
the optimal estimator. Although they are unknown in practice, it is
clear that Condition C3 is fairly mild.

For Condition C1, we need to make use of the orthogonality and
boundedness of certain basis functions to properly design $h(\bX)$
in the constraints. Suppose $Z=0, 1, 2$ and the empirical CDF of the
one dimensional auxiliary covariate $X$ is
$F_n(x)=n^{-1}\sum_{i=1}^{n} 1_{\{X_i\leq x\}}$. By making use of
multivariate Fourier expansion, the arguments can be generalized to
the high dimensional auxiliary covariate case.  Let
$\bg_{r_n}^*(\bbeta)=(\bm^T(\bbeta;Y,Z),(1_1-\pi_1),\widehat{s}_{11},\widehat{c}_{11},
\ldots,\widehat{s}_{1d_n},\widehat{c}_{1d_n},(1_2-\pi_2),\widehat{s}_{21},\widehat{c}_{21},
\ldots,\widehat{s}_{2d_n},\widehat{c}_{2d_n})^T$, where
$1_k=1_{(Z=k)}$, $r_n=4d_n+q+2$,
$\widehat{s}_{ij}=(1_i-\pi_i)\sin(2\pi j F_n(X))$,
$\widehat{c}_{ij}=(1_i-\pi_i)\cos(2\pi j F_n(X))$, $i=1,2$,
$j=1,\ldots,d_n$. It can be shown that, when $d_n=o(n^{1/4})$,
(i)-(iii) are satisfied. For example, we can apply the fact that
those basis functions are orthogonal when their arguments are
$U[0,1]$ and they are bounded to show (i) and (ii) hold.
Because the procedure is invariant under linear transformations, the
eigenvalues can grow with $n$ if all of them grow at the same rate.
However, we do not believe in general they can grow at different
rates since the covariance matrix is sandwiched in the
variance-covariance expression, which needs to be well-conditioned.
Furthermore, (iii) can be verified by taking the expansion of the
conditional expectations. Likewise, we may apply other orthogonal
basis functions that are bounded. For example, we can use the
Legendre polynomials of $(2F_n(X)-1)$ which are bounded by 1 on
[-1,1]. Legendre polynomials, i.e. $1, x, (3x^2-1)/2,\ldots$, are
linear transformations of polynomial terms $1, x, x^2,\ldots$.
Therefore we can also use polynomial terms of $(2F_n(X)-1)$ in the
auxiliary constraints due to linear transformation invariance of the
empirical likelihood.  As pointed out by a referee, the standard
independence assumption for empirical likelihood is violated due to
the plug-in estimator $F_n$.  Intuitively, the validity of using
$F_n$ instead of $F$ relies on the fact that those constraints are
still zero-mean conditioning on all the covariates. A rigorous proof
can be found in the Appendix.

Analogous to the case with a fixed number of constraints, let
$l(\bbeta)=\sum_{i=1}^{n}\log\Big(1+\widehat{\blambda}_n^T(\bbeta)\bg_{n,i}(\bbeta)\Big)$.
The empirical likelihood ratio statistic for testing $H_0: \bbeta =
\bbeta_0$ is \vspace{-3mm}\begin{equation} T_{1n} =
2l(\bbeta_0)-2l(\widehat{\bbeta}_n).\vspace{-3mm}
\end{equation}
Then under Conditions C1-C3, the Wilks type theorem of convergence
to the $\chi^2$ distribution is still valid for testing the null
hypothesis of no treatment effect. \vspace{-2mm}
\begin{thm}\label{thm2}
Suppose that Conditions C1-C3 are satisfied. Then, under the null
hypothesis $H_0$,  $T_{1n}$ converges in distribution to
$\chi_{(q)}^2$ as $n\To \infty$.\vspace{-2mm}\end{thm}

More generally, we can test hypothesis on a subset of treatment
effects $\bbeta$ instead of all components of it. For instance, we
may be interested in testing whether $\bbeta_2=0$ in the simple
example (\ref{mean2gr}). Specifically, let $\bbeta^T=(\bbeta_1^T,
\bbeta_2^T)^T$, where $\bbeta_1$ and $\bbeta_2$ are $q_1$- and
$q_2$-vectors, respectively.  For $\widetilde{H_0}: \bbeta_1 =
\bbeta_{10}$, the profile empirical likelihood ratio test statistic
is simply\vspace{-3mm}
\begin{equation}\label{T2n}
T_{2n} =
2l(\bbeta_{10},\widehat{\bbeta}_{20})-2l(\widehat{\bbeta}_n),\vspace{-3mm}
\end{equation}
where $\widehat{\bbeta}_{20}$ minimizes $l(\bbeta_{10},\bbeta_2)$
with respect to $\bbeta_2$. The following result shows that a
Wilks type $\chi^2$  approximation still holds.\vspace{-3mm}
\begin{cor}\label{cor1} Suppose that Conditions C1-C3 are satisfied.
Then, under the null hypothesis, $T_{2n}$ converges in
distribution to $\chi_{(q_1)}^2$ as $n\To
\infty$.\vspace{-2mm}\end{cor}

Auxiliary information can be used to not only increase the
precision of estimated treatment effects, but to also increase
power in hypothesis testing. To evaluate power, we need to derive
the asymptotic distribution of the test statistic under the
alternative hypothesis.  We shall consider the contiguous
alternative which deviates from the null by the order of
$O(n^{-1/2})$; cf. Hajek, Sidak and Sen (1999) and Serfling
(1980). For notational convenience, let $\bA =
\bD_m^T\widetilde{\bSigma}^{-1}\bD_m$ and write
\begin{equation*} \bA
= \begin{bmatrix}
\bA_{11}& \bA_{12}\\
\bA_{21}& \bA_{22}\end{bmatrix},
\end{equation*}
where $\bA_{ij}=E(\partial
\bm^T(\bbeta_0)/\partial\bbeta_i)\widetilde{\bSigma}^{-1}E(\partial
\bm(\bbeta_0)/\partial\bbeta_j^T)$, $i=1,2$ and $j=1,2$.

\begin{thm}\label{thm3} Suppose that Conditions C1-C3 are satisfied.
Then under the sequence of contiguous alternatives $H_a:
\bbeta=\bbeta_a=\bbeta_0+\bh/\sqrt{n}$, the empirical likelihood
ratio test statistic $T_{1n}$ converges in distribution to a
noncentral $\chi^2$ with degrees of freedom $q$ and noncentrality
parameter $\bh^TA\bh$.\end{thm} \vspace{-3mm} Similarly, the
noncentrality parameter of the limiting $\chi^2$ distribution
becomes the projected Fisher information when there are nuisance
parameters. \vspace{-3mm}
\begin{cor}\label{cor2} Under the same assumptions as those in
Theorem \ref{thm3} and with $H_a$ replaced by $\widetilde{H_a}:
\bbeta_1=\bbeta_{1a}=\bbeta_{10}+\bh_1/\sqrt{n}$, the empirical
likelihood ratio test statistic $T_{2n}$ in (\ref{T2n}) converges
in distribution to a noncentral $\chi^2$ with degrees of freedom
$q_1$ and noncentrality parameter
$\bh_1^T(\bA_{11}-\bA_{12}\bA^{-1}_{22}\bA_{21})\bh_1$.\end{cor}

It can be seen that the empirical likelihood approach reproduces the
standard asymptotic results in parametric likelihood theory (Cox and
Hinkley 1974).  Similar to the estimation problem, adding more
constraints will result in more powerful tests. When the number of
constraints goes to infinity, the corresponding tests become
asymptotically most powerful.

\section{NUMERICAL STUDIES}

In this section, we discuss computational issues arising from
implementing the constrained optimization problems and report
simulation results associated with the empirical likelihood based
covariate adjustment method.

The primary step in computing the empirical likelihood is to
maximize (\ref{EL}) subject to constraints (\ref{constraints}). The
lagrangian is
$$\mathbb{P}_\star(p, \bbeta, \blambda,
\gamma)=\sum_{i=1}^n \log_\star(p_i)+n\blambda^T\sum_{i=1}^n
p_i\bg_r(\bbeta;y_i,z_i,\bx_i)+n\gamma(\sum_{i=1}^n p_i - 1),$$
where $\blambda$ and $\gamma$ are the Lagrange multipliers and
$\log_\star$ is a modified natural logarithm defined in Owen
(2001). Thus, we obtain estimators for $p$ and $\bbeta$ by
differentiating $\mathbb{P}_\star$ with respect to $p$, $\bbeta$,
$\blambda$ and $\gamma$ and setting them to $0$.

Working directly with $n+q+r+1$ free variables involves gradient and
Hessian matrices of daunting dimensions.  Alternatively we may use
the two-stage Newton algorithm as discussed in Section 3.1 that can
eliminate some parameters. Nonetheless, unlike the usual testing
case where $\bbeta$ is fixed at $\bbeta_0$, the outer stage in the
two-stage Newton algorithm, i.e. minimization over $\bbeta$ while
keeping $\blambda$ fixed, is difficult in practice because of the
possibility of a non-positive definite Hessian matrix. Zedlewski
(2008) points out that ``Concentrating out some parameters leads to
a smaller optimization problem, but it can make it more difficult.
Thus the two-stage Newton algorithm is fast but unreliable and can
lead to frustrating convergence problems.  In most cases $n$ is much
greater than $q+r$, so the largest block of the Hessian is an
$n\times n$ diagonal matrix.''.  In our implementation, we use a
Matlab package ``matElike'', which solves the primal problem by
including modern optimization codes exploiting matrix sparsity. We
find the package to be both robust and fast.  The link to the Matlab
package and the code to implement our method can be found at
\url{http://www.stat.columbia.edu/~xwu/software.html}.\\

\noindent 4.1 \textbf{Estimation}\\
\noindent The simulation results reported below are all based on
$5000$ Monte Carlo replications.  The sample size is chosen to be
$200$ throughout.  We consider the case of two treatment groups
with the treatment indicator $Z$ generated with
$P(Z=0)=P(Z=1)=0.5$. The response variable $Y$ is binary with
$\logit\{E(Y|Z)\}=\beta_1+\beta_2Z$.  The parameter of interest is
either $\bbeta=(\beta_1, \beta_2)^T$ or $\beta_2$.

In the first scenario, the auxiliary covariate $X$ is generated as
a one dimensional Normal random variable with mean 0 and different
variances.  The magnitude of the variance correlates with the
influence of X on the response. Given $Z$ and $X$, $Y$ is then
generated as Bernoulli according to
$\logit\{P(Y=1|Z=g,X)\}=\alpha_{0g}+\alpha_gX$, where
$\alpha_{00}=0.3, \alpha_{01}=1, \alpha_0=1, \alpha_1=1.5$ and $g=0$
or $1$.

 From Table~\ref{table1} we see that when the standard deviation of X is 2,
the Monte Carlo standard errors gradually decrease and approach
the optimal ones.  From ``marginal'' to ``5 Fourier'', the standard
errors drop significantly. However, additional constraints beyond
``5 Fourier'' do not appear to have much impact on further variance
reduction. Note that a large number of additional constraints
require substantially more computing time.  Thus, we will only
compare the results of ``marginal'' with ``5 Fourier'' in the other
cases.  A single (i.e., nonparallel) process that calculates the
maximum empirical likelihood estimate and the p-value for testing
the null hypothesis of no treatment difference takes, on average,
less that 2 seconds to run for a data set of $200$ samples using $5$
constraints. The computation time is estimated using a 2.33GHz
processor on a server with 8GB RAM.

Table~\ref{table1} also shows that the means of Monte Carlo
estimates differ from the true value of $\bbeta$ at the third
decimal place and the coverage probabilities are around 0.95.  The
Monte Carlo standard errors of estimates from five estimating
equations are generally smaller than those from marginal models. The
improvement becomes more pronounced when the variance of X becomes
larger. Also, the average length of 95\% Wald confidence intervals
are smaller than those of marginal models.

In the second scenario, the link function is quadratic in X, i.e.,
$\logit\{P(Y=1|Z=g,X)\}=\alpha_{0g}+\alpha_gX^2$, with the same
$\alpha_0g$ and $\alpha_g$ values, $g=0, 1$. From
Table~\ref{table2}, we see that the coverage probabilities are
satisfactory and close to their nominal levels as in the first
scenario.  The biases are slightly larger, however, they are still
small relative to the standard errors. As expected, the Monte Carlo
standard errors and the average lengths of 95\% Wald confidence
intervals from five estimating equations are smaller than those from
the two marginal ones.

In the third scenario, there are two auxiliary covariates $X_1$ and
$X_2$ and the response Y is generated as
$\logit\{P(Y=1|Z=g,\bX)\}=\alpha_{0g}+\alpha_{1g}X_1+\alpha_{2g}X_2$,
g=0,1, with $\alpha_{00}=0.3, \alpha_{01}=1, \alpha_{10}=1,
\alpha_{11}=1.5, \alpha_{20}=2, \alpha_{21}=1.5$.  The estimating
equations for the marginal method remain the same since there is no
covariate adjustment involved.  Let $\kappa(Z)=\sqrt{2}(2Z-1)$ and
$W_k=2\pi F_n(X_k)$, $k=1, 2$.  The empirical likelihood method with
constraints, $\kappa(Z)$, $\kappa(Z)\sin(W_1)$,
$\kappa(Z)\cos(W_1)$, $\kappa(Z)\sin(W_2)$, $\kappa(Z)\cos(W_2)$,
except the marginal estimating equations is denoted by ``7
Fourier''.  From Table~\ref{table3}, the performance of the
estimates is similar to the previous two
scenarios.\\

\noindent 4.2 \textbf{Testing}\\
\noindent With the same data generating process as in the preceding
subsection, the corresponding hypothesis testing results are
presented in Tables~\ref{table4}, ~\ref{table5} and ~\ref{table6}.
In each scenario, the profile empirical likelihood ratio test is
used to test the null hypothesis $\widetilde{H_0}: \beta_2 = 0$.
CovProb denotes coverage probabilities for testing
$\beta_2=\beta_{20}$. We have the following observations.  First, in
all three tables, both coverage probabilities of the profile
empirical likelihood ratio tests are close to the nominal 95\%
level. Second, the attained power from 5 estimating equations is
larger than that from marginal estimating equations. Third, when X
is one dimensional, the gain in power is more significant as the
standard deviation of X increases.

\section{APPLICATION}

We apply the proposed empirical likelihood based approach to the
Global Use of Strategies to Open Occluded Coronary Arteries
(GUSTO)-I trial data, which were kindly provided to us by Karen Pieper from
the Duke Clinical research Institute. The primary endpoint was
30-day death, which occurred in 6.29\% of 10366 patients randomly
assigned to tissue plasminogen activator (TPA) (g=1), 7.32\% of
10354 patients randomly assigned to skreptokinase (SK) with IV
heparin (g=2), 6.99\% of 10303 patients randomly assigned to a combination
of SK and TPA (g=3) and 7.24\% of 9773 patients randomly assigned to
SK with SQ heparin (g=4). Besides treatment assignment and outcome,
some baseline auxiliary covariates concerning demographics (age,
sex, weight, height), risk factors (hypertension, diabetes, smoking,
hypercholesterolemia), other history (family history of MI, previous
MI, previous angina, previous revascularization) and presenting
characteristics (blood pressure, tachycardia, anterior infarct
location, killip class, ST elevation on electrocardiography) were
recorded on each subject. In Steyerberg et al. (2000), the relative
prognostic strength of 17 baseline covariates was evaluated by their
univariate $\chi^2$ model, which was calculated as the difference in
-2 log-likelihood between a univariate logistic regression model
with and without the characteristic.  The strongest prognostic
factor was age and this was further confirmed by the $R^2$ measure on
the log-likelihood scale, which approximately indicated the
percentage of variance explained. Except for the calculation of
correlation, adjustment for important predictors such as age is
always recommended in the case of short-term death after acute
myocardial infarction. Thus, we will compare
unadjusted and age-adjusted results for the four treatment
groups.

The marginal model between the 30-day death (Y) and treatment
assignment (Z) is given by $\logit\{E(Y|Z)\}=\beta_1+\beta_2
1_{(Z=2)}+\beta_3 1_{(Z=3)}+\beta_4 1_{(Z=4)}$. For the age(X)
adjustment, we use 9 auxiliary constraints $(1_{(Z=g)}-0.25)$,
$(1_{(Z=g)}-0.25)F_n(x)$ and $(1_{(Z=g)}-0.25)F_n^2(x)$, $g=2,3,4$,
where $F_n(x)$ is the empirical c.d.f. of age.

The unadjusted estimates
$(\widehat{\beta}_1,\widehat{\beta}_2,\widehat{\beta}_3,\widehat{\beta}_4)$
are (-2.7014, 0.1630, 0.1129, 0.1517) with standard errors (0.04109,
0.05619, 0.05670, 0.05557). Estimates adjusted for age are (-2.7014,
0.1628, 0.1126, 0.1521) with standard errors (0.04109, 0.05619,
0.05670, 0.05556).  The p-values for the unadjusted and adjusted
hypothesis testing of $\beta_2=\beta_3=\beta_4=0$ are 0.0136 and
0.0135, respectively.

The unadjusted test is already significant, so the additional
improvement in $p$-value after covariate adjustment only reconfirms
the scientific conclusion. However, if the sample size were smaller,
the change in $p$-value might be more consequential. For
illustrative purposes, we randomly draw a subsample of size 20000
from the complete data and pretend that is what we had in reality.
In one of these cases, the $p$-values for the unadjusted test of
$\beta_2=\beta_3=\beta_4=0$ is 0.0391 while it becomes 0.0362 after
adjusting for age.  In another case, it changes from 0.0508 to
0.0458. \vspace{-8mm}
\section{DISCUSSION}
Nonparametric covariate adjustment is of importance in analysis of
randomized clinical trial data. When properly done, it can result in
efficiency improvement while maintaining the nonparametric nature of
the usual tests. Empirical likelihood approach is nonparametric,
constraint based and efficient in extracting information from data.

For randomized clinical trials, covariate information with no model
assumption can be extracted from certain type of constraints or
estimating equations. We propose an empirical likelihood based
approach for covariate adjustment. The resulting likelihood ratio
test is shown to have the usual Wilks type $\chi^2$ approximation,
with increased power as the number of constraints increases. The
corresponding maximum empirical likelihood estimate also enjoys
similar asymptotic properties. We demonstrate that the $\chi^2$ and
normal approximations continue to hold as the number of constraints
grows with sample size. We further show that in doing so the
semiparametric efficiency can be achieved.

One of the practical issues is how to select basis functions in the
constraints. From our experiences with simulations and real data
analysis, it appears that there is no universal way to deal with
this issue. A related issue is how many basis functions should be
used. One ad hoc way to do that is to consider variance reduction
when additional constraints are added. We believe that if initial
basis functions are properly chosen, then only a very small number
of constraints will be needed.

It will be of interest to extend this empirical likelihood based
nonparametric covariate adjustment to other situations, including
observational studies. Of particular
importance are the survival and longitudinal studies where the
response variables may be dependent or causal. For survival data, Lu
and Tsiatis (2008) have introduced a general model framework for
covariate adjustment and derived a semiparametric efficient score.
We believe a similar approach, which makes use of suitable covariate
based constraints and achieves the asymptotic efficiency, can be
developed.

\section*{APPENDIX}
Here we provide proofs of the theoretical results presented in the
previous sections. For notational convenience, let
$G_n(\bbeta)=\underset{\tiny1\leq i\leq n}
{\max}\norm{\bg_{n,i}(\bbeta)}$,\quad
$\bSigma_{n,m}=E(\bm_{r_n}(\bbeta_0)\bm_{r_n}^T(\bbeta_0))$,\quad
$\bSigma_{n,opt}=E(\bm_{r_n}^{opt}(\bbeta_0)(\bm_{r_n}^{opt}(\bbeta_0))^T)$,\quad
$\bD_{r_n}=E(\partial\bg_{r_n}(\bbeta_0)/\partial\bbeta^T)$,\quad
$\bD_{m_n}=E(\partial\bm_{r_n}(\bbeta_0)/\partial\bbeta^T)$, and
$\bD_{opt}=E(\partial\bm_{r_n}^{opt}(\bbeta_0)/\partial\bbeta^T)$.

\begin{lem}\label{convexhull}
The probability that zero is outside the convex hull spanned by
$\{\bg_{n,i}, i=1,\ldots, n\}$ goes to zero as
$n\to\infty$.\vspace{-5mm}
\end{lem}
Proof.  This follows from Lemma 4.2 in Hjort et al. (2009) and
discussions thereof.\qed

\begin{lem}\label{lem1}
Under (i),(ii) and C2, the eigenvalues of $\bS_n(\bbeta_0)$ are
bounded away from 0 and $\infty$.
\end{lem}
Proof.  It can be shown by making use of proofs of condition (D4)
and Lemma 4.5 in Hjort et al. (2009).\qed

\begin{lem}\label{lem2}
Under (i),(ii) and C2,\begin{eqnarray}
\norm{\widehat{\blambda}_n(\bbeta_0)}&=&O_p(n^{-1/2}r_n^{1/2})\label{lem31}\\
\underset{\tiny\norm{\bbeta-\bbeta_0}\leq
n^{-1/3}}\sup\norm{\widehat{\blambda}_n(\bbeta)}&=&O_p(n^{-1/3})\label{lem32}\\
\underset{\tiny\norm{\bbeta-\bbeta_0}\leq
n^{-1/3}}\sup\norm{\widehat{\blambda}_n(\bbeta)-\bS_n(\bbeta)^{-1}
\overline{\bg}_n(\bbeta)}&=&O_p(n^{-2/3}r_n^{1/2}).\label{lem33}\end{eqnarray}
\end{lem}

Proof.  Under (i),(ii) and C2, we can apply results in Portnoy
(1988) to get\begin{equation}\label{portnoy}
\norm{n^{1/2}\overline{\bg}_n(\bbeta_0)}=O_p(r_n^{1/2}).\end{equation}
Under (i),\begin{equation}\label{Gn} G_n(\bbeta) \leq
Mr_n^{1/2}=O_p(r_n^{1/2}).\end{equation} Write
$\widehat{\blambda}_n(\bbeta)=\norm{\widehat{\blambda}_n(\bbeta)}\bu_n(\bbeta)$,
where $\norm{\bu_n(\bbeta)}=1$.  Then similar to (\ref{lam}), we can
show that
$$
0 =
\bu_n^T(\bbeta)\frac{1}{n}\sum_{i=1}^{n}\frac{\bg_{n,i}(\bbeta)}
{1+\widehat{\blambda}_n^T(\bbeta)\bg_{n,i}(\bbeta)}\leq
 \bu_n^T(\bbeta)\overline{\bg}_n(\bbeta)-\frac{\norm{\widehat{\blambda}_n
 (\bbeta)}}{1+\norm{\widehat{\blambda}_n(\bbeta)}G_n(\bbeta)}mineig(\bS_n(\bbeta)),$$
where $mineig(\bM)$ stands for the minimum eigenvalue of the matrix
$\bM$. Therefore, we have
\begin{equation}\label{eqn1}
\norm{\widehat{\blambda}_n(\bbeta)}(mineig(\bS_n(\bbeta))-\bu_n^T(\bbeta)
\overline{\bg}_n(\bbeta)G_n(\bbeta))\leq
\bu_n^T(\bbeta)\overline{\bg}_n(\bbeta),\end{equation} from which we
know that (\ref{lem31}) holds due to (\ref{portnoy}), (\ref{Gn}) and
Lemma~\ref{lem1}.

When $\norm{\bbeta-\bbeta_0}\leq n^{-1/3}$, define\vspace{-5mm}
\begin{equation}
L_n =
\max_{j,k}\abs{\bS_{n,j,k}(\bbeta)-\bS_{n,j,k}(\bbeta_0)}.\vspace{-5mm}
\end{equation}
Using the same technique as in Lemma~\ref{lem1}, $r_nL_n = o_p(1)$
ensures that the minimum eigenvalue of $S_n(\bbeta)$ is bounded away
from zero. Since there are only finitely many terms in $\bg_{r_n}$
containing $\bbeta$, due to the $\delta$-method, this can be further
reduced to $\norm{\bbeta-\bbeta_0}=o(r_n^{-1})$, which is true under
C2. By expanding $\overline{\bg}_n(\bbeta)$ in the $n^{-1/3}$
neighborhood of $\bbeta_0$, we obtain $
\overline{\bg}_n(\bbeta)=O_p(n^{-1/3})$ uniformly in
$\norm{\bbeta-\bbeta_0}\leq n^{-1/3}$.  Then (\ref{lem32}) follows
from equation (\ref{eqn1}).

We know that $\widehat{\blambda}_n(\bbeta)$ satisfies the constraint
$n^{-1}\sum_{i=1}^{n}\bg_{n,i}(\bbeta)/\{1+\widehat{\blambda}_n^T(\bbeta)\bg_{n,i}(\bbeta)\}=0,$
which implies
\begin{equation}\label{lamexp}
\widehat{\blambda}_n(\bbeta)=\bS_n(\bbeta)^{-1}\overline{\bg}_n(\bbeta)+
\bS_n(\bbeta)^{-1}\frac{1}{n}\sum_{i=1}^{n}\bg_{n,i}(\bbeta)\frac{\bu_n^T
(\bbeta)\bg_{n,i}(\bbeta)\bg_{n,i}^T(\bbeta)\bu_n(\bbeta)}{1+\widehat{\blambda}_n^T
(\bbeta)\bg_{n,i}(\bbeta)}\norm{\widehat{\blambda}_n(\bbeta)}^2.
\end{equation}
By the triangle inequality and some simple algebra, the final term
in (\ref{lamexp}) is bounded by $O_p(n^{-2/3}r_n^{1/2})$. Since
$\norm{\bS_n(\bbeta)^{-1}}=O_p(1)$, (\ref{lem33}) follows from
(\ref{lamexp}).\qed \vspace{-3mm}
\begin{lem}\label{lem3}
Under Conditions C1-C3,
$$\norm{\bD_{r_n}^T\bSigma_{n,g}^{-1}\bD_{r_n} - \bD_m^T\widetilde{\bSigma}^{-1}\bD_m} = o(1).$$
\end{lem}
Proof.  Let $\bm_{r_n}=\bm(\bbeta;Y,Z)+\bA_n\bW_n\bV_{n}$. Since
$\bA_n\bW_n\bV_{n}$ does not involve $\bbeta$, we
have\begin{equation}\label{nobeta}\bD_{m_n}^T\bSigma_{n,m}^{-1}\bD_{m_n}=\bD_m^T\bSigma_{n,m}^{-1}\bD_m,\end{equation}
which by (iii), converges to $\bD_m^T\widetilde{\bSigma}^{-1}\bD_m.$

Second, following Qin and Lawless (1994), for any $n$, we have
\begin{equation*}
\bD_{r_n}^T\bSigma_{n,g}^{-1}\bD_{r_n}
=\bD_{opt}^T\bSigma_{n,opt}^{-1}\bD_{opt},
\end{equation*}
where $\bm_{r_n}^{opt}=\bA_{opt}(\bbeta)\bg_{r_n}$ is a $q$-vector
and $\bA_{opt}(\bbeta)$ is the optimal linear combination of
$\bg_{r_n}$.  So it suffices to show the following difference is
zero:\begin{equation}\label{vardiff}
\bD_{opt}^T\bSigma_{n,opt}^{-1}\bD_{opt}-\bD_m^T\bSigma_{n,m}^{-1}\bD_m.\end{equation}
Given
 (\ref{nobeta}), (\ref{vardiff}) is positive definite due to optimality. Furthermore,
\begin{eqnarray*}
&&\bD_{opt}^T\bSigma_{n,opt}^{-1}\bD_{opt}-\bD_m^T\bSigma_{n,m}^{-1}\bD_m\\
&=&
\bD_{opt}^T\bSigma_{n,opt}^{-1}\bD_{opt}-\bD_m^T\widetilde{\bSigma}^{-1}\bD_m+\bD_m^T\widetilde{\bSigma}^{-1}\bD_m-\bD_m^T\bSigma_{n,m}^{-1}\bD_m.
\end{eqnarray*}
By Zhang et al. (2008), we know that
$\bD_m^T\widetilde{\bSigma}^{-1}\bD_m$ is the semiparametric
efficiency bound, which implies the first difference is non-positive
definite. Since the second difference is $o_p(1)$, we know
(\ref{vardiff}) is nonpositive definite.  \qed \vspace{-3mm}
\begin{lem}\label{lem4}
Under (i), (ii) and C2, $\norm{\widehat{\bbeta}_n-\bbeta_0} <
n^{-1/3}$.
\end{lem}
Proof. We first consider $\bbeta$ on the $n^{-1/3}$ sphere of
$\bbeta_0$, i.e. $\bbeta-\bbeta_0=\bu n^{-1/3}$, where $\bu$ is a
unit vector.  On the one hand, by the Taylor series expansion and
Lemma~\ref{lem2},
\begin{equation*}
 2\sum_{i=1}^{n}\log\Big(1+\widehat{\blambda}_n^T(\bbeta)\bg_{n,i}(\bbeta)\Big)\\
 =2n\widehat{\blambda}_n^T(\bbeta)\overline{\bg}_n(\bbeta)-n\widehat{\blambda}_n^T
 (\bbeta)\bS_n(\bbeta)\widehat{\blambda}_n(\bbeta)+O_p(r_n^{1/2}).\end{equation*}
By (\ref{lem33}), it is equivalent to
$n\overline{\bg}_n^T(\bbeta)\bS_n^{-1}(\bbeta)\overline{\bg}_n(\bbeta)+O_p(r_n^{1/2}).$
By taking the Taylor series expansion at $\bbeta_0$, it equals to
$$\bu^T\bD_{r_n}^T\bSigma_{n,g}^{-1}\bD_{r_n}\bu n^{1/3}+o_p(n^{1/3}),$$ which is bounded below by
 $O_p(n^{1/3})$ by
 Lemma~\ref{lem3}.  On the other hand,
$2\sum_{i=1}^{n}\log\Big(1+\widehat{\blambda}_n^T(\bbeta_0)\bg_{n,i}(\bbeta_0)\Big)
= O_p(r_n),$ which is strictly less than $O_p(n^{1/3})$ by condition
C2.  Therefore, $\norm{\widehat{\bbeta}_n-\bbeta_0} < n^{-1/3}$.
\qed

\begin{lem}\label{lem5}
Under conditions C1-C3, we have the asymptotic normality of the
``influence function''
$$\bD_{r_n}^T\bSigma_{n,g}^{-1}n^{1/2}\overline{\bg}_n(\bbeta_0)\To
N(0, \bD_m^T\widetilde{\bSigma}^{-1}\bD_m).$$
\end{lem}

Proof. We can reduce the problem to the unidimensional case by
noting that it suffices to show that for any $q\times1$ vector
$\bt$,\begin{equation}\label{uni}\bt^T\bD_{r_n}^T\bSigma_{n,g}^{-1}n^{1/2}\overline{\bg}_n(\bbeta_0)\To
N(0, \bt^T\bD_m^T\widetilde{\bSigma}^{-1}\bD_m\bt).\end{equation}
First, the variance of the left hand side of (\ref{uni}) is
$\bt^T\bD_{r_n}^T\bSigma_{n,g}^{-1}\bD_{r_n}\bt$, which converges to
$\bt^T\bD_m\widetilde{\bSigma}^{-1}\bD_m\bt$ by Lemma~\ref{lem3}.

Second, we verify the Lindeberg condition (Billingsley 1986)
\begin{eqnarray*}
&&\sum_{i=1}^{n}E\Big\{\Big[n^{-1/2}\bt^T\bD_{r_n}^T\bSigma_{n,g}^{-1}\bg_{n,i}(\bbeta_0)\Big]^2
\mathbf{1}_{\Big[\abs{n^{-1/2}\bt^T\bD_{r_n}^T\bSigma_{n,g}^{-1}\bg_{n,i}(\bbeta_0)}>\varepsilon
\Big]}\Big\}\\
&=&E\Big\{\Big[\bt^T\bD_{r_n}^T\bSigma_{n,g}^{-1}\bg_{r_n}(\bbeta_0)\Big]^2
\mathbf{1}_{\Big[\abs{\bt^T\bD_{r_n}^T\bSigma_{n,g}^{-1}\bg_{r_n}(\bbeta_0)}>n^{1/2}\varepsilon
\Big]}\Big\}\To 0,
\end{eqnarray*}
where the last step comes from
$$P\bigg(\abs{\bt^T\bD_{r_n}^T\bSigma_{n,g}^{-1}\bg_{r_n}(\bbeta_0)}>n^{1/2}\varepsilon
\bigg) \leq
E\bigg(\bt^T\bD_{r_n}^T\bSigma_{n,g}^{-1}\bg_{r_n}(\bbeta_0)\bigg)^2
\bigg/n\varepsilon^2,$$ which goes to $0$ since the numerator is
asymptotically bounded. Hence Lemma~\ref{lem5} holds by the
Lindeberg-Feller Central Limit
Theorem.\qed\\

Proof of Theorem~\ref{thm1}.  Let
$\bU_n(\bbeta,\blambda)=\frac{1}{n}\sum_{i=1}^{n}\frac{\bg_{n,i}(\tiny\bbeta)}{1+\tiny\blambda^T\bg_{n,i}(\tiny\bbeta)}$
and
$\bV_n(\bbeta,\blambda)=\frac{1}{n}\sum_{i=1}^{n}\frac{\tiny\blambda
\partial
\bg_{n,i}^T(\tiny\bbeta)/\partial\tiny\bbeta}{1+\tiny\blambda^T\bg_{n,i}(\tiny\bbeta)}.$
We know that $(\widehat{\bbeta}_n,\widehat{\blambda}_n)$ satisfies
$\bU_n(\widehat{\bbeta}_n,\widehat{\blambda}_n)=0$ and
$\bV_n(\widehat{\bbeta}_n,\widehat{\blambda}_n)=0$.  By taking the
Taylor series expansion, we have
\begin{eqnarray}
0&=&\bU_n(\widehat{\bbeta}_n,\widehat{\blambda}_n)\nonumber\\
&=&\overline{\bg}_n(\bbeta_0)+\hat{\bD}^T(\bbeta_0)
(\widehat{\bbeta}_n-\bbeta_0)-\bS_n(\bbeta_0)\widehat{\blambda}_n+O_p(n^{-2/3}r_n^{1/2})
\label{U},\quad \text{and}\\
0&=&\bV_n(\widehat{\bbeta}_n,\widehat{\blambda}_n)\nonumber\\
&=&\hat{\bD}^T(\bbeta_0)\widehat{\blambda}_n+
O_p(n^{-2/3})\label{V}.
\end{eqnarray}
Solving (\ref{U}) and (\ref{V}) for $\widehat{\bbeta}_n-\bbeta_0$,
we get,
\begin{equation}\label{betahat}
n^{1/2}(\widehat{\bbeta}_n-\bbeta_0)=-n^{1/2}(\hat{\bD}(\bbeta_0)^T\bS_n^{-1}(\bbeta_0)
\hat{\bD}(\bbeta_0))^{-1}
\hat{\bD}(\bbeta_0)\bS_n^{-1}(\bbeta_0)\overline{\bg}_n(\bbeta_0)+o_p(1).
\end{equation}
By triangular inequality and Lemma~\ref{lem3}, we can show that
\begin{equation}\label{inverse}\norm{(\hat{\bD}^T(\bbeta_0)\bS_n^{-1}(\bbeta_0)
\hat{\bD}(\bbeta_0))^{-1}-(\bD_m^T\widetilde{\bSigma}^{-1}\bD_m)^{-1}}
= o_p(1).
\end{equation}
By Lemma~\ref{lem5},
\begin{eqnarray*}
\hat{\bD}^T(\bbeta_0)\bS_n^{-1}(\bbeta_0)n^{1/2}\overline{\bg}_n(\bbeta_0)&=&
\bD_{r_n}^T\bSigma_{n,g}^{-1}n^{1/2}\overline{\bg}_n
(\bbeta_0)+o_p(n^{-1/2+\varepsilon}
r_n^{1/2})\\
&\To& N(0, \bD_m^T\widetilde{\bSigma}^{-1}\bD_m).
\end{eqnarray*}
Then Theorem~\ref{thm1} follows from (\ref{betahat}),
(\ref{inverse}) and Slutsky's Theorem.
\qed\\

Proof of Theorem~\ref{varestimate}.  Since there are only finitely
many terms in $\overline{\bg}_n$ and $\bS_n$ that contain $\bbeta$,
by the $\delta$-method, we have
$$\norm{(\hat{\bD}^T(\widehat{\bbeta}_n)\bS_n^{-1}(\widehat{\bbeta}_n)\hat{\bD}(\widehat{\bbeta}_n))^{-1}-(\hat{\bD}^T(\bbeta_0)\bS_n^{-1}(\bbeta_0)\hat{\bD}(\bbeta_0))^{-1}}=o_p(1).$$
Then
the result follows from (\ref{inverse}). \qed\\

Proof of Theorem~\ref{thm2}. Taking the Taylor series expansion, we
get
\begin{eqnarray*}
T_{1n} &=& n^{1/2}(\widehat{\bbeta}_n-\bbeta_0)^T
\Big[\frac{\partial^2}{\partial\bbeta\partial\bbeta^T}\frac{1}{n}\sum_{i=1}^{n}\log
\Big(1+\widehat{\blambda}_n^T(\bbeta_0)\bg_{n,i}(\bbeta_0)\Big)\Big]n^{1/2}
(\widehat{\bbeta}_n-\bbeta_0)+o_p(1)\\
&=& n^{1/2}(\widehat{\bbeta}_n-\bbeta_0)^T \bA
n^{1/2}(\widehat{\bbeta}_n-\bbeta_0)+o_p(1).
\end{eqnarray*}
Then Theorem~\ref{thm1} implies $T_{1n}\To \chi_{q}^2$ as $n\To
\infty$, when $H_0$ is true.
\qed\\

Proof of Corollary~\ref{cor1}.  When only $\bbeta_1$ is specified in
the null hypothesis, we write the likelihood ratio statistic as
the sum of two differences, each of which can be expanded in a manner similar
to that in Theorem~\ref{thm2} and we have
\begin{eqnarray*}
T_{2n} &=&
\Big[2\sum_{i=1}^{n}\log\Big(1+\widehat{\blambda}_n^T(\bbeta_0)\bg_{n,i}(\bbeta_0)\Big)
-2\sum_{i=1}^{n}\log\Big(1+\widehat{\blambda}_n^T(\widehat{\bbeta}_n)\bg_{n,i}
(\widehat{\bbeta}_n)\Big)\Big]\\
&&-\Big[2\sum_{i=1}^{n}\log\Big(1+\widehat{\blambda}_n^T(\bbeta_0)\bg_{n,i}(\bbeta_0)\Big)
-2\sum_{i=1}^{n}\log\Big(1+\widehat{\blambda}_n^T(\bbeta_{10},\widehat{\bbeta}_{20})
\bg_{n,i}(\bbeta_{10}, \widehat{\bbeta}_{20})\Big)\Big]\\
&=& n^{1/2}(\bbeta_{10}-\widehat{\bbeta}_{1n})^T(\bA_{11}-\bA_{12}
\bA_{22}^{-1}\bA_{21})n^{1/2}(\bbeta_{10}-\widehat{\bbeta}_{1n})+o_p(1).
\end{eqnarray*}
The last equation comes from
$\widehat{\bbeta}_{20}-\bbeta_{20}=\widehat{\bbeta}_{2n}-\bbeta_{20}+\bA_{22}^{-1}
\bA_{21}(\widehat{\bbeta}_{1n}-\bbeta_{10})+o_p(1)$. Thus
Corollary~\ref{cor1} holds because
$n^{1/2}(\widehat{\bbeta}_{1n}-\bbeta_{10})$ converges in
distribution to $N(0,
(\bA_{11}-\bA_{12}\bA_{22}^{-1}\bA_{21})^{-1})$ under
$\widetilde{H_0}$.
\qed\\

Proof of Theorem~\ref{thm3}.  Following the same steps as in the
proof of Theorem~\ref{thm1}, we can show that
$$n^{1/2}\bA^{-1/2}(\widehat{\bbeta}_n-\bbeta_0)\To
N(\bA^{-1/2}\bh, \bI).$$ Taking the Taylor series expansion of the
empirical likelihood ratio test statistic at $\bbeta_0$, we have
\begin{equation*} T_{1n}=
n^{1/2}(\widehat{\bbeta}_n-\bbeta_a+\bh/\sqrt{n})^T\bA(\bbeta_0)n^{1/2}
(\widehat{\bbeta}_n-\bbeta_a+\bh/\sqrt{n})+o_p(1),
\end{equation*}
where the second equality comes from
$\bbeta_a=\bbeta_0+\bh/\sqrt{n}$ being a sequence of contiguous
alternatives. Therefore, $T_{1n}\To \chi_{q}^2$ with noncentrality
parameter $\bh^T\bA\bh$ as $n\To \infty$ under the alternative
$H_a: \bbeta=\bbeta_a=\bbeta_0+\bh/\sqrt{n}$.
\qed\\

Proof of Corollary~\ref{cor2}. Similar to the preceding proof, we
have under the contiguous alternative
$$
T_{2n} =
n^{1/2}(\bbeta_{10}-\widehat{\bbeta}_{1n})^T(\bA_{11}-\bA_{12}
\bA_{22}^{-1}\bA_{21})n^{1/2}(\bbeta_{10}-\widehat{\bbeta}_{1n})+o_p(1).
$$
Similar to Theorem~\ref{thm1}, we can show that when
$\widetilde{H_a}: \bbeta_1=\bbeta_{1a}=\bbeta_{10}+\bh_1/\sqrt{n}$
is true, $(\bA_{11}-\bA_{12}\bA_{22}^{-1}\bA_{21})^{-1/2}n^{1/2}
(\bbeta_{10}-\widehat{\bbeta}_{1n})$ converges in distribution to
$N((\bA_{11}-\bA_{12}\bA_{22}^{-1}\bA_{21})^{-1/2}\bh_1,
\bI),$ which implies Corollary~\ref{cor2}.\qed\\

In the following part of the APPENDIX, we verify that $\bg_{r_n}^*$
in the examples following Corollary~\ref{cor3} satisfies Condition
C1. The other conditions are satisfied trivially. Since the Fourier
basis are naturally bounded by $1$, the uniform boundedness reduces
to the boundedness of $\bm$ which is of finite dimension and usually
holds easily.  So (i) is satisfied. Let
$$\bV_n=((1_1-\pi_1)/\pi_1,s_{11},c_{11},
\ldots,s_{1d_n},c_{1d_n},(1_2-\pi_2)/\pi_2,s_{21},c_{21},
\ldots,s_{2d_n},c_{2d_n})^T$$ and
$\bg_{r_n}(\bbeta)=(\bm^T(\bbeta;Y,Z), \bV^T_{n})^T$, where
$1_k=1_{(Z=k)}$, $s_{ij}=\sqrt{2}(1_i-\pi_i)\sin(2\pi j
F(X))/\pi_i$, $c_{ij}=\sqrt{2}(1_i-\pi_i)\cos(2\pi j F(X))/\pi_i$,
$i=1,2$, $j=1,\ldots,d_n$.  For notation simplicity, we omit $\bW_n$
in $\bW_n\bV_n$ when there is no ambiguity.  Then letting $\bI_d$
denote the $d\times d$ identity matrix, we have the following matrix
partition
\[\bSigma_{n,g}=\left[\begin{array}{c|c|c}
E(\bm\bm^T)&\multicolumn{2}{c}{E(\bm\bV_{n}^T)}\\\hline
&\frac{1-\pi_1}{\pi_1}\bI_{2d_n+1}&-\bI_{2d_n+1}\\\cline{2-3}
E(\bV_{n}\bm^T)&-\bI_{2d_n+1}&\frac{1-\pi_2}{\pi_2}\bI_{2d_n+1}
\end{array}\right].\]
Thus, by some simple algebra and C3, we can show that the
eigenvalues of $\bSigma_{n,g}$ are bounded away from 0 and $\infty$.
However, since F is unknown in practice, we typically use
$F_n(x)=n^{-1}\sum_{i=1}^{n} 1_{\{X_i\leq x\}}$ instead. Let
$$\widehat{\bV}^T_{n}(z,x)=((1_1-\pi_1)/\pi_1,\widehat{s}_{11},\widehat{c}_{11},
\ldots,\widehat{s}_{1d_n},\widehat{c}_{1d_n},(1_2-\pi_2)/\pi_2,\widehat{s}_{21},\widehat{c}_{21},
\ldots,\widehat{s}_{2d_n},\widehat{c}_{2d_n})$$ and
$\widehat{\bg}_{r_n}(\bbeta)=(\bm^T(\bbeta;Y,Z),
\widehat{\bV}^T_{n}(Z,X))^T$, where
$\widehat{s}_{ij}=\sqrt{2}(1_i-\pi_i)\sin(2\pi j F_n(x))/\pi_i$,
$\widehat{c}_{ij}=\sqrt{2}(1_i-\pi_i)\cos(2\pi j F_n(x))/\pi_i$,
$i=1,2$, $j=1,\ldots,d_n$. Define $\bvarepsilon_n =
\widehat{\bg}_{r_n}\widehat{\bg}_{r_n}^T-\bg_{r_n}\bg_{r_n}^T$. Then
\begin{eqnarray*}
r_n\underset{\tiny j,k}\max\abs{\bvarepsilon_{n,j,k}} &\leq&
2M^2r_n\abs{\sin\pi
d_n(F_n(X)-F(X))}\\
&=& O_p(r_n^2n^{-1/2}).
\end{eqnarray*}
Following the argument in Lemma~\ref{lem1}, when we let
$r_n=o(n^{\frac{1}{4}})$, we know the eigenvalues of
$E(\widehat{\bg}_{r_n}(\bbeta_0)\widehat{\bg}_{r_n}^T(\bbeta_0))$
are also bounded away from zero and infinity.  So (ii) holds.

Moreover, let
$f(z,x)=\sum_{k=0}^K(1_k-\pi_k)E(\bm(\bbeta;Y,Z)|Z=k,x)$ and $\bA_n$
be the Fourier coefficients in the Fourier expansion of $f(z,x)$
with the Fourier basis specified in $\widehat{\bV}_n(z,x)$. We know
from Fourier approximation theory that $\bA_n\widehat{\bV}_n(z,x)\To
f(z,x)$ uniformly.  Thus, by Condition C3 and the Dominated
Convergence
Theorem, (iii) is satisfied.\qed\\

Proof of the validity of the plug-in estimator $F_n$. Checking the
derivation of all the theorems, we find that the following two
conditions will guarantee the validity of the theorems when $F$ is
replaced by $F_n$
\begin{eqnarray}
\norm{n^{-1/2}\sum_{i=1}^n (\hat{\bg}_{n,i}-\bg_{n,i})} = o_p(1)\label{aim3}\\
\norm{n^{-1}\sum_{i=1}^n
\Big\{\hat{\bg}_{n,i}\hat{\bg}_{n,i}^T-\bg_{n,i}\bg_{n,i}^T\Big\}}
= o_p(1)\label{aim4},
\end{eqnarray}
where $\bg_{n,i}$ and $\hat{\bg}_{n,i}$ are $\bg_{r_n}$ and
$\hat{\bg}_{r_n}$ evaluated at the $i^{th}$ sample.  The norm of a
matrix $\bM$ is defined to be $\underset{\bu}{\sup}\norm{\bM\bu}$,
where $\bu$ is a unit vector.  The sufficiency of the above two
conditions when the number of constraints is fixed can be seen from
the existing literature (see, for example, Hjort et al. (2009)).

Denote the $j^{th}$ component of a vector $\bg$ by $\bg^j$.  Then,
for any $j$, we have
$$E\bigg[\bigg\{n^{-1/2}\sum_{i=1}^n(\hat{\bg}^j_{n,i}-\bg^j_{n,i})\bigg\}^2
\bigg|\bX_1,\ldots,\bX_n\bigg]\leq
C_1r^2_n\norm{F_n-F}^2_{\infty},$$ where $C_1$ is a universal
constant. Therefore,
$$E\bigg\{\norm{n^{-1/2}\sum_{i=1}^n(\hat{\bg}_{n,i}-\bg_{n,i})}^2\bigg|\bX_1,\ldots,\bX_n\bigg\}\leq
C_1r^3_n\norm{F_n-F}^2_{\infty}=O_p(r^3_n/n),$$ which converges to
$0$ in probability due to C2.  By Chebyshev's inequality, we know
that for any $\varepsilon>0$,
$$P\bigg\{\norm{n^{-1/2}\sum_{i=1}^n(\hat{\bg}_{n,i}-\bg_{n,i})}\geq\varepsilon
\bigg|\bX_1,\ldots,\bX_n\bigg\}=o_p(1),$$
which implies (\ref{aim3}) due to the dominated convergence theorem.

Denote $\bvarepsilon_u = n^{-1}\sum_{i=1}^n
\Big\{\hat{\bg}_{n,i}\hat{\bg}_{n,i}^T-\bg_{n,i}\bg_{n,i}^T\Big\}\bu$.
Then we have $E\{(\bvarepsilon^j_u)^2|\bX_1,\ldots,\bX_n\}\leq
O_p(r_n^3n^{-2})$ uniformly for $\bu$ and $j$. Therefore,
$E\{\norm{\bvarepsilon_u}^2|\bX_1,\ldots,\bX_n\}\leq
O_p(r_n^4n^{-2})\leq o_p(1)$, which implies (\ref{aim4}).\qed

\newpage
\vspace{10mm}

\begin{table}
\caption{Bias and Standard Error Comparisons When Logit is Linear in
X.}
\begin{center}
    \begin{tabular}{ccccccc}
    \hline
    Method& True $\bbeta$&MC Bias&OptStd&MC Std&CovProb&avlen\\\hline
    \\
    \multicolumn{7}{c}{$X \sim N(0,0.5^2)$}\\
    marginal &0.2832&0.0033&0.1992&0.2025&0.9520&0.7960\\
             &0.6096&0.0063&0.2872&0.3007&0.9486&1.1801\\
    5 Fourier&0.2832&0.0036&0.1992&0.2027&0.9500&0.7870\\
             &0.6096&0.0056&0.2872&0.2968&0.9468&1.1536\\
    \\
             \multicolumn{7}{c}{$X \sim N(0,1^2)$}\\
    marginal &0.2479&0.0010&0.1929&0.2025&0.9520&0.7940\\
             &0.4634&0.0063&0.2585&0.2988&0.9472&1.1562\\
    5 Fourier&0.2479&0.0011&0.1929&0.1992&0.9496&0.7718\\
             &0.4634&0.0049&0.2585&0.2812&0.9424&1.0785\\
        \\
            \multicolumn{7}{c}{$X \sim N(0,2^2)$}\\
     marginal &0.1814& 0.0040&0.1800&0.1995&0.9526&0.7912\\
              &0.2792& 0.0003&0.2110&0.2951&0.9452&1.1324\\
    5 Fourier &0.1814& 0.0043&0.1800&0.1873&0.9518&0.7337\\
              &0.2792&-0.0018&0.2110&0.2439&0.9418&0.9292\\
    7 Fourier &0.1814&0.0030&0.1800&0.1860&0.9494&0.7186\\
              &0.2792&0.0008&0.2110&0.2341&0.9442&0.8846\\
    9 Fourier &0.1814&0.0032&0.1800&0.1857&0.9464&0.7101\\
              &0.2792&0.0008&0.2110&0.2311&0.9384&0.8631\\
    11 Fourier&0.1814&0.0032&0.1800&0.1852&0.9448&0.7037\\
              &0.2792&0.0007&0.2110&0.2293&0.9340&0.8490\\
    \hline
    \end{tabular}
    \label{table1}
    \end{center}
\end{table}

\vspace{5mm}

\noindent NOTE: In all the tables, `marginal' means using empirical
likelihood method with 2 marginal estimating equations
$Y-\phi(\beta_1+\beta_2Z)$ and
$Z\Big(Y-\phi(\beta_1+\beta_2Z)\Big)$, while ``5 Fourier'' has three
additional estimating equations $2Z-1$, $\sqrt{2}(2Z-1)\sin[2\pi
F_n(X)]$ and $\sqrt{2}(2Z-1)\cos[2\pi F_n(X)]$, where $F_n(X)$ is
the empirical cumulative distribution function of X.  MC Bias is
Monte Carlo bias, OptStd is the asymptotic standard error obtained
according to the sandwich formula, MC Std is the Monte Carlo
standard error, CovProb is the coverage probability of 95\% Wald
confidence intervals and avlen is the average length of those
confidence intervals.

\newpage

\vspace{10mm}

\begin{table}
\caption{Bias and Standard Error Comparisons When Logit is Quadratic
in X.}
\begin{center}
    \begin{tabular}{ccccccc}
    \hline
    Method& True $\bbeta$&MC Bias&OptStd&MC Std&CovProb&avlen\\\hline
    \\
    \multicolumn{7}{c}{$X \sim N(0,0.5^2)$}\\
    marginal&0.5298&0.0057&0.2059&0.2093&0.9516&0.8160\\
            &0.7758&0.0094&0.3169&0.3266&0.9536&1.2683\\
    5 Fourier&0.5298&0.0061&0.2059&0.2088&0.9480&0.8090\\
             &0.7758&0.0088&0.3169&0.3257&0.9524&1.2523\\
    \\
             \multicolumn{7}{c}{$X \sim N(0,1^2)$}\\
     marginal&0.9664&0.0106&0.2182&0.2307&0.9476&0.8845\\
             &0.8105&0.0182&0.3466&0.3795&0.9494&1.4450\\
    5 Fourier&0.9664&0.0111&0.2182&0.2254&0.9448&0.8604\\
             &0.8105&0.0156&0.3466&0.3648&0.9502&1.3800\\
    \hline
    \end{tabular}
    \label{table2}
    \end{center}
\end{table}

\vspace{3cm}

\begin{table}
\caption{Bias and Standard Error Comparisons When Logit Contains Two
Covariates.}
\begin{center}
    \begin{tabular}{ccccccc}
    \hline
    Method& True $\bbeta$&MC Bias&OptStd&MC Std&CovProb&avlen\\\hline
    \\
    \multicolumn{7}{c}{$X_1 \sim N(0,1^2), X_2 \sim N(0,2^2)$}\\
    marginal &0.1061&-0.0003&0.1649&0.2005&0.9558&0.7883\\
             &0.3157& 0.0043&0.1828&0.2933&0.9494&1.1282\\
    7 Fourier&0.1061&-0.0009&0.1649&0.1761&0.9526&0.6813\\
             &0.3157& 0.0051&0.1828&0.2311&0.9438&0.8716\\
    \\
    \multicolumn{7}{c}{$X_1 \sim N^2(0,1^2), X_2 \sim N(0,2^2)$}\\
    marginal &0.4389&0.0063&0.1688&0.2032&0.9550&0.8069\\
             &0.5493&0.0056&0.1985&0.3072&0.9486&1.2023\\
    7 Fourier&0.4389&0.0052&0.1688&0.1825&0.9494&0.7012\\
             &0.5493&0.0041&0.1985&0.2490&0.9458&0.9396\\
    \\
    \multicolumn{7}{c}{$X_1 \sim N^2(0,0.5^2), X_2 \sim N^2(0,1^2)$}\\
     marginal&1.4746& 0.0149&0.2482&0.2594&0.9562&1.0201\\
             &0.5813& 0.0233&0.3857&0.4310&0.9498&1.6363\\
    7 Fourier&1.4746& 0.0144&0.2482&0.2512&0.9518&0.9771\\
             &0.5813& 0.0224&0.3857&0.4126&0.9486&1.5485\\
    \hline
    \end{tabular}
    \label{table3}
    \end{center}
\end{table}

\vspace{3mm}

\noindent NOTE: The logit is either quadratic ($X\sim
N^2(\cdot,\cdot)$) or linear ($X\sim N(\cdot,\cdot)$) in each
covariate.

\newpage

\begin{table}
\caption{Power Comparison When Logit is Linear in X.}
\begin{center}
\begin{tabular}{ccccccc}
\hline
      &         &           & \multicolumn{2}{c}{marginal} & \multicolumn{2}{c}{5 Fourier}\\\cline{4-7}
X & $\beta_{10}$ & $\beta_{20}$ & CovProb & Power & CovProb & Power\\\hline\\
$N(0,0.5^2)$& 0.2125& 0.8304& 0.9498& 0.7928& 0.9492& 0.8216\\
$N(0,1^2)$  & 0.1379& 0.8207& 0.9494& 0.7826& 0.9458& 0.8682\\
$N(0,2^2)$  & 0.0386& 0.8182& 0.9486& 0.7938& 0.9436& 0.9568\\
\hline
\end{tabular}
\label{table4}
\end{center}
\end{table}

\begin{table}
\caption{Power Comparison When Logit is Quadratic in X.}
\begin{center}
\begin{tabular}{ccccccc}
\hline
      &         &           & \multicolumn{2}{c}{marginal} & \multicolumn{2}{c}{5 Fourier}\\\cline{4-7}
X & $\beta_{10}$ & $\beta_{20}$ & CovProb & Power & CovProb & Power\\\hline\\
$N(0,0.5^2)$& 0.8511& 1.0599& 0.9442& 0.8428& 0.9448& 0.8498\\
$N(0,1^2)$  & 0.9662& 0.9359& 0.9464& 0.7356& 0.9482& 0.7724\\
\hline
\end{tabular}
\label{table5}
\end{center}
\end{table}

\begin{table}
\caption{Power Comparison When Logit Contains Two Covariates.}
\begin{center}
\begin{tabular}{ccccccc}
\hline
      &         &           & \multicolumn{2}{c}{marginal} & \multicolumn{2}{c}{7 Fourier}\\\cline{4-7}
$X_1$, $X_2$ & $\beta_{10}$ & $\beta_{20}$ & CovProb & Power & CovProb & Power\\\hline\\
$N(0,1^2)$, $N(0,2^2)$& 0.0694& 0.8461& 0.9488& 0.8166& 0.9430& 0.9308\\
$N^2(0,1^2)$, $N(0,2^2)$  & 0.2468& 0.7012& 0.9418& 0.6584& 0.9438& 0.8636\\
$N^2(0,0.5^2)$, $N^2(0,1^2)$& 1.1701& 0.8342& 0.9496& 0.5873& 0.9478& 0.6140\\
\hline
\end{tabular}
\label{table6}
\end{center}
\end{table}

\vspace{3mm}

\noindent NOTE: In each scenario, $\beta_{10}$ and $\beta_{20}$ are
the true values.  The profile empirical likelihood ratio test is
used to test the null hypothesis $\widetilde{H_0}: \beta_2 = 0$.
CovProb are the coverage probabilities of tests
$\beta_2=\beta_{20}$.

\end{document}